\documentclass[twocolumn,superscriptaddress,showpacs,preprintnumbers,amsmath,amssymb,oneside,prl]{revtex4}
\usepackage{graphicx}
\usepackage{dcolumn}
\usepackage{bm}

\begin{document}

%%%%%%%%%%%%%%%%%%%%%%%%%%%%%%%%%%%%%%%%%%%%%%%%%%%%%%%%%%%%%%%%%%%

\title{Optical Nondestructive Controlled-NOT Gate without Using Entangled Photons}

\author{Xiao-Hui Bao}
\affiliation{Hefei National Laboratory for Physical Sciences at
Microscale and Department of Modern Physics, University of Science
and Technology of China, Hefei, Anhui 230026, China}

\author{Teng-Yun Chen}
\affiliation{Hefei National Laboratory for Physical Sciences at
Microscale and Department of Modern Physics, University of Science
and Technology of China, Hefei, Anhui 230026, China}

\author{Qiang Zhang}
\affiliation{Hefei National Laboratory for Physical Sciences at
Microscale and Department of Modern Physics, University of Science
and Technology of China, Hefei, Anhui 230026, China}

\author{Jian Yang}
\affiliation{Hefei National Laboratory for Physical Sciences at
Microscale and Department of Modern Physics, University of Science
and Technology of China, Hefei, Anhui 230026, China}

\author{Han Zhang}
\affiliation{Hefei National Laboratory for Physical Sciences at
Microscale and Department of Modern Physics, University of Science
and Technology of China, Hefei, Anhui 230026, China}

\author{Tao Yang}
\affiliation{Hefei National Laboratory for Physical Sciences at
Microscale and Department of Modern Physics, University of Science
and Technology of China, Hefei, Anhui 230026, China}

\author{Jian-Wei Pan}
\affiliation{Hefei National Laboratory for Physical Sciences at
Microscale and Department of Modern Physics, University of Science
and Technology of China, Hefei, Anhui 230026, China}
\affiliation{Physikalisches Institut der Universitaet Heidelberg,
Philosophenweg 12, Heidelberg 69120, Germany}

\date{\today}

%%%%%%%%%%%%%%%%%%%%%%%%%%% abstract %%%%%%%%%%%%%%%%%%%%%%%%%%%%%%%

\begin{abstract}
We present and experimentally demonstrate a novel optical
nondestructive controlled-NOT gate without using entangled ancilla.
With much fewer measurements compared with quantum process
tomography, we get a good estimation of the gate fidelity. The
result shows a great improvement compared with previous experiments.
Moreover, we also show that quantum parallelism is achieved in our
gate and the performance of the gate can not be reproduced by local
operations and classical communications.
\end{abstract}

\pacs{03.67.Lx, 03.65.Ud, 42.25.Hz}

\maketitle

%%%%%%%%%%%%%%%%%%%%%%% introduction %%%%%%%%%%%%%%%%%%%%%%%%%%%%%%
The controlled-NOT (CNOT) or similar entangling gates between two
individual quantum bits (qubits) are essential for quantum
computation \cite{Nielsen2000,KNILL2001}. Also entangling gates can
be utilized to construct a complete Bell-state analyzer which is
required in various quantum communication protocols
\cite{bennett1992,bennett1993,bennett1996}. Photons are one of the
best candidates for qubit due to the robustness against decoherence
and ease of single-qubit operation. So far there have been several
experiments implementing the optical CNOT gate
\cite{pittman2003,brien2003,Langford2005,nikolai2005,ryo2005,zhao2005,sara2004}.
These experiments can be divided into two groups, one is the
destructive CNOT gate
\cite{pittman2003,brien2003,Langford2005,nikolai2005,ryo2005} which
means that one have to measure the output of the gate to verify a
successful operation, imposing a great limitation for its further
implementations, and the other is the nondestructive gate
\cite{zhao2005,sara2004}.

For a nondestructive CNOT gate, the information whether the
operation succeeds or not is provided. This information can then be
utilized for future conditional operations on the photonic qubits to
achieve efficient linear optical quantum computation. Also with this
information arbitrary entangled state can be constructed in an
efficient way, especially the cluster state for one-way quantum
computation \cite{Raussendorf2001,duan2005}. So nondestructive CNOT
gate is much more important than the destructive one. To build a
nondestructive gate, usually ancilla photons are unavoidably
required. Previous scheme \cite{pittman2001} requires an entangled
photon pair as assistance. The well developed SPDC (spontaneous
parametric down-conversion) \cite{kwiat1995} entangled photon source
will be unsuitable due to the probabilistic character. Generating
entangled photons directly from quantum dots
\cite{STEVENSON:2006,AKOPIAN:2006} is still at its beginning and the
fidelity is to be improved. Making use of entangled photons
generated from single photons \cite{wang2003,BROWNE2005,ZHANG2006}
is another solution, but it will make the setup much more
complicated and reduce the success probability a lot under the
present technology. Also the imperfections of the entangled photon
pair will cause a degradation to the fidelity of the gate, making
high-precision gate operation even more difficult to achieve.

%%%%%%%%%%%%%%%%%%%%%%% the scheme %%%%%%%%%%%%%%%%%%%%%%%%%%%%%%%%%

Here we present and experimentally demonstrate a novel
nondestructive CNOT gate without using entangled photons but only
single photons instead, which is a great improvement compared with
former scheme \cite{pittman2001}. With much fewer measurements
compared with quantum process tomography \cite{poyatos1997}, we get
a good estimation of the gate fidelity with the method developed by
Hofmann in \cite{hofmann2005, hofann2005pra}.

\begin{figure}[hbtp]
\includegraphics[width=6cm]{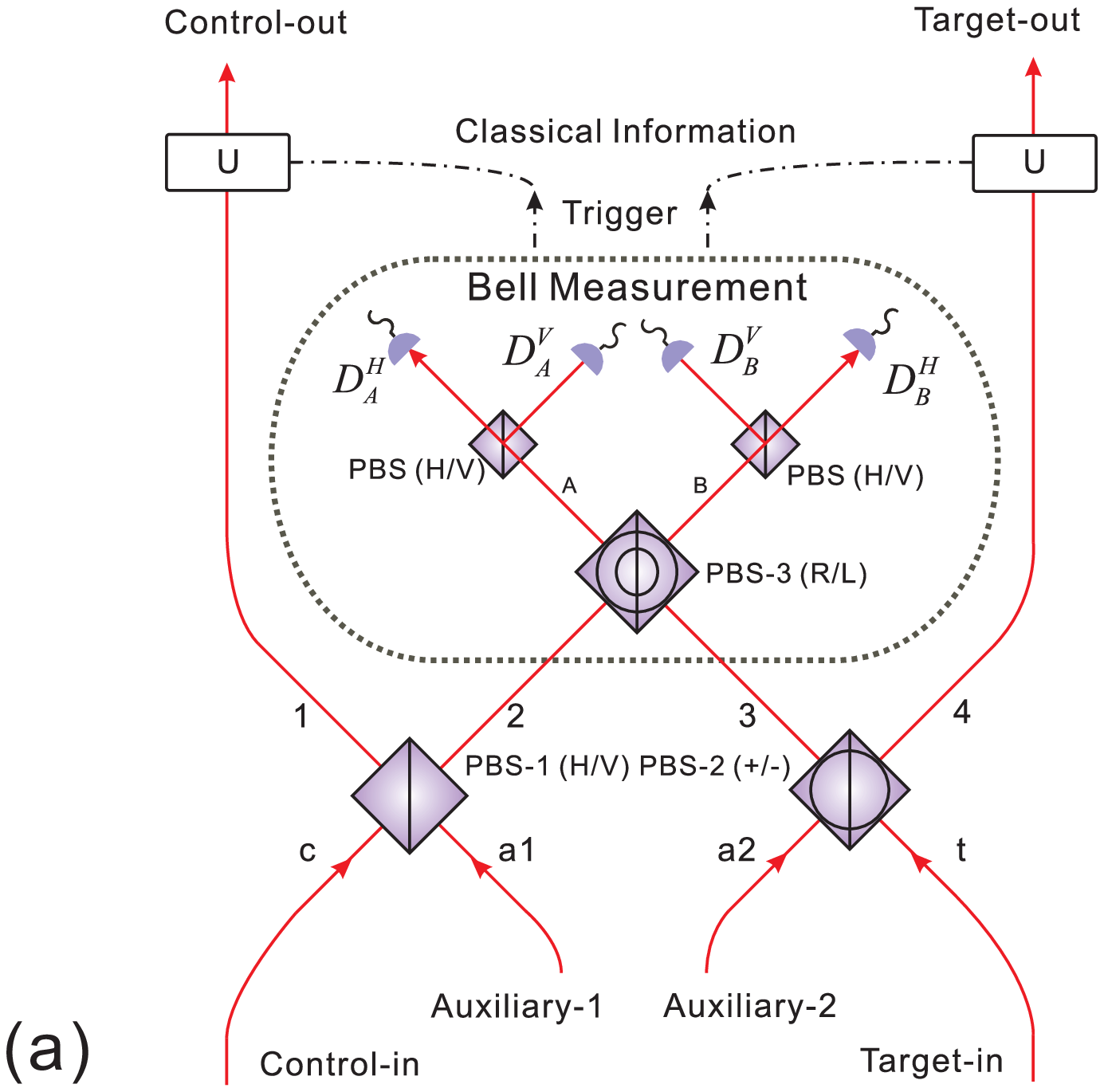}
\includegraphics[width=6cm]{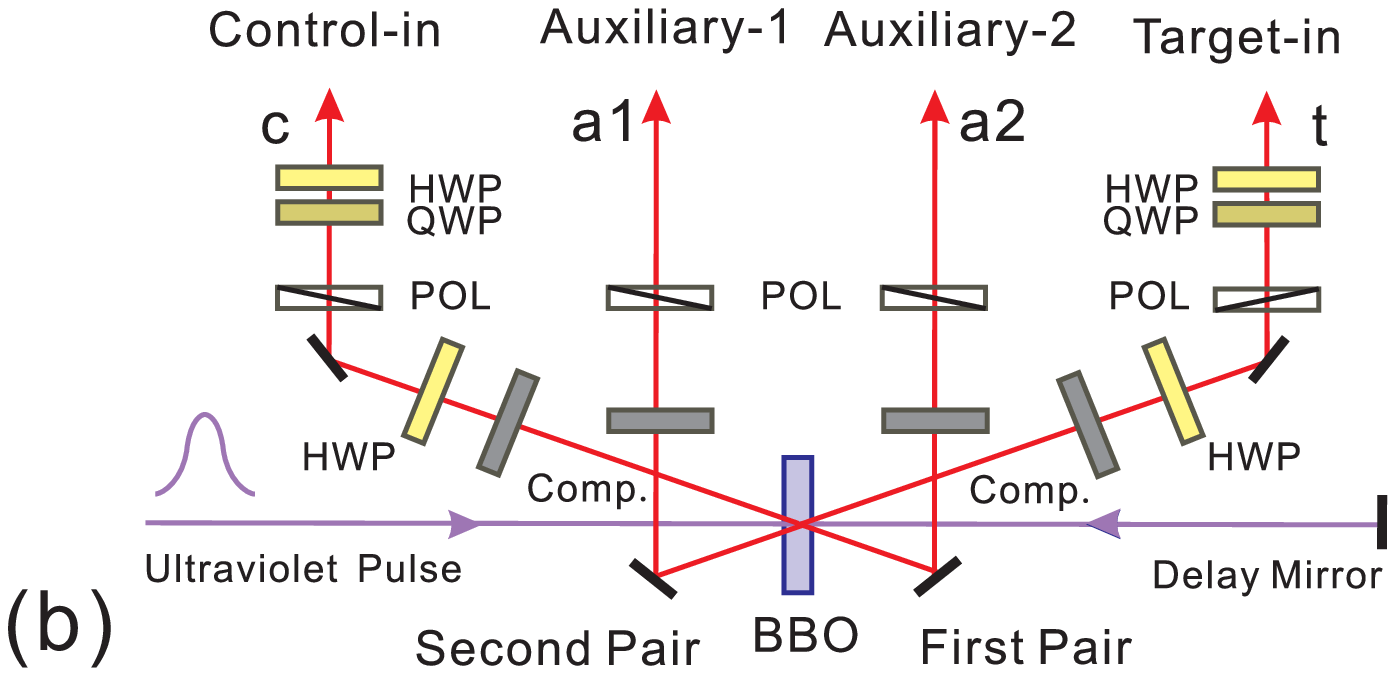}\\
\caption{(color online). (a) Our scheme to implement nondestructive
CNOT gate with polarization beam splitters (PBS) in
$|H\rangle$/$|V\rangle$ basis, in $|+\rangle$/$|-\rangle$ basis and
in $|R\rangle$/$|L\rangle$ basis. PBS in $|+\rangle$/$|-\rangle$
basis (PBS-2) is constructed with a PBS in $|H\rangle$/$|V\rangle$
basis and four half-wave plates (HWP); PBS in
$|R\rangle$/$|L\rangle$ basis (PBS-3) is constructed with a PBS in
$|H\rangle$/$|V\rangle$ basis and four quarter-wave plates (QWP).
This gate works like this: four photons (control qubit, target qubit
and two auxiliary qubit) enter from the bottom; if there is a
coincident count between detector $D_A$ and detector $D_B$, a
successful CNOT gate operation will be made after sending one bit
classical information and doing the corresponding single-qubit
unitary operations on photon 1 and 4. Then the state of photon 1 is
exactly the output of the control qubit; and the state of photon 4
is exactly the output of the target qubit. In our proof-of-principle
experiment, for simplification only the coincident events between
$D_A^H$ and $D_B^H$ are registered, and a HWP is added to do the
corresponding $\sigma_z$ operation on photon 1. (b) Experimental
setup to generate the required four photons. Near infrared
femtosecond laser pulses ($\approx$200 fs, 76 MHz, 788 nm) are
converted to ultraviolet pulses through a frequency doubler LBO
($LiB_3O_5$) crystal (not shown). Then the ultraviolet pulse
transmits through the main BBO ($\beta-BaB_2O_4$) crystal (2mm)
generating the first photon pair, then reflected back generating the
second photon pair. Compensators (Comp.) which is composed of a
HWP($45^\circ$) and a BBO crystal (1mm) are added in each arm. The
observed 2-fold coincident count rate is about $1.2\times10^4/s$. In
each arm we add a polarizer to do the disentanglement and set the
initial product four-photon state to
$|H\rangle_c|+\rangle_{a1}|H\rangle_{a2}|H\rangle_t$. Additional
wave plates are added in path c and path t to prepare arbitrary
polarization states.}\label{fig1}
\end{figure}

In our scheme, the qubit we consider refers to the polarization
state of photons. We define the polarization state $|H\rangle$ as
logic 0 and $|V\rangle$ as logic 1. Let's assume that the input sate
of the control qubit is $|\psi\rangle^c=
\alpha|H\rangle+\beta|V\rangle$ and of the target qubit is
$|\psi\rangle^t= \gamma|H\rangle+\delta|V\rangle$. As shown in Fig.
\ref{fig1}(a), two auxiliary photons with polarization state of
$1/\sqrt{2}(|H\rangle+|V\rangle)$ and $|H\rangle$ are required. Then
the total state of the input four photons can be expressed as:
\begin{eqnarray}
|\Psi\rangle_1 & = &
\frac{1}{2}(\alpha|H\rangle_c+\beta|V\rangle_c)\
(|H\rangle_{a1}+|V\rangle_{a1}) \nonumber
\\
& & \quad \,(|+\rangle_{a2}+|-\rangle_{a2})\
(\gamma|H\rangle_t+\delta|V\rangle_t)
\end{eqnarray}
where the subscript (c, a1 ,a2 and t) represents each path of the
four photons, state $|+\rangle$ equals to
$1/\sqrt{2}(|H\rangle+|V\rangle)$ and $|-\rangle$ equals to
$1/\sqrt{2}(|H\rangle-|V\rangle)$. First the four photons transmit
through PBS-1 which transmits state $|H\rangle$ and reflects state
$|V\rangle$ and PBS-2 which transmits state $|+\rangle$ and reflects
state $|-\rangle$. Let's consider the case that there is one photon
in each output path. Then the four-photon state will change to
\begin{eqnarray}
|\Psi\rangle_2 & = &
[\alpha|H\rangle_1|H\rangle_{2}+\beta|V\rangle_1|V\rangle_{2}]\otimes
\nonumber
\\
& &
[(\gamma+\delta)|+\rangle_{3}|+\rangle_4+(\gamma-\delta)|-\rangle_{3}|-\rangle_4]
\end{eqnarray}
with a probability of 1/4. This state expanded in the Bell basis of
photon 2 and photon 3 is shown as follows:
\begin{equation}
\begin{array}{ccrc}
|\Psi\rangle_2 & = & I_1I_4U_{14} & |\psi\rangle^c_{1} \ |\psi\rangle^t_{4}\otimes|\Phi^+\rangle_{23}\\
& + & I_1\sigma_{x4}U_{14} & |\psi\rangle^c_{1} \ |\psi\rangle^t_{4}\otimes|\Psi^+\rangle_{23}\\
\ & + & \sigma_{z1}I_4U_{14} & |\psi\rangle^c_{1} \ |\psi\rangle^t_{4}\otimes|\Phi^-\rangle_{23}\\
& + & \sigma_{z1}\sigma_{x4}U_{14} & |\psi\rangle^c_{1} \
|\psi\rangle^t_{4}\otimes|\Psi^-\rangle_{23}
\end{array}
\label{expand}
\end{equation}
where $U$ refers to the CNOT operation; $|\Phi^\pm\rangle$ and
$|\Psi^\pm\rangle$ are standard Bell states in $|H\rangle/|V\rangle$
basis; $\sigma_x$ and $\sigma_z$ are Pauli operators with the form
$\sigma_x=|H\rangle\langle V|+|V\rangle\langle H|$,
$\sigma_z=|H\rangle\langle H|-|V\rangle\langle V|$.

From Equ. \ref{expand} we can see that if the jointly measured
result of photon 2 and photon 3 is the state $|\Phi^+\rangle$, then
the state of photon 1 and photon 4 is exactly the output state of
the CNOT operation; if the measured result is other state
($|\Phi^-\rangle$, $|\Psi^+\rangle$ or $|\Psi^-\rangle$), then
corresponding single qubit operations on the state of photon 1 and
photon 4 are required to get the result of the CNOT operation. But
within the linear optical technology only two of the four bell
states can be distinguished. In our scheme as shown in Fig.
\ref{fig1}(a), the two Bell states are $|\Phi^-\rangle$ and
$|\Psi^+\rangle$. $|\Phi^-\rangle$ corresponds to the coincidence
between $D_A^H$ and $D_B^H$ or between $D_A^V$ and $D_B^V$ and
$|\Psi^+\rangle$ corresponds to the coincidence between $D_A^H$ and
$D_B^V$ or between $D_A^V$ and $D_B^H$. In conclusion if there is a
coincident count between $D_A$ ($D_A^H$ or $D_A^V$) and $D_B$
($D_B^H$ or $D_B^V$), then one bit of classical information will be
sent to do the corresponding single qubit operation as shown in Fig.
\ref{fig1}(a), and after that the state of photon 1 and photon 4
will be the exact output state of the CNOT operation. The total
success probability is $1/8$.

For each PBS (PBS-1 and PBS-2) the output can be divided into three
cases: one in each output path; two in first path and zero in the
second; zero in the first and two in the second. Consider PBS-1 and
PBS-2 jointly, there will be nine cases as follows:
\begin{equation*}
\begin{array}{cl}
Group \ 1 & 1:1:1:1 \\
Group \ 2 & 1:1:2:0 \quad 1:1:0:2 \quad 2:0:1:1 \\
& 0:2:1:1 \quad 2:0:0:2 \quad 0:2:2:0 \\
Group \ 3 & 2:0:2:0 \quad 0:2:0:2 \\
\end{array}
\end{equation*}
where \textit{$n_1:n_2:n_3:n_4$} corresponds to the photon numbers
in each path (1, 2, 3 or 4). Group 1 is what we expected, just as
what we have discussed. In group 2 the total number of photons on
the path 2 and path 3 does not equal to 2, so the cases in this
group will not give a correct trigger signal with assistance of
photon number resolving detectors \cite{Fujiwara2005}. For the cases
in group 3, the total photon number of path 2 and path 3 equals to
2. Roughly thinking, these two cases will lead to a coincidence
between $D_A$ and $D_B$, which will ruin this scheme. But
considering the photon bunching effect \cite{Eisenberg2005}, we will
find that it is not possible for a correct trigger signal, because
two photons either one in $|H\rangle$ and the other in $|V\rangle$
or one in $|+\rangle$ and the other in $|-\rangle$ will go to the
same output path when they pass through a PBS in the R/L basis
(PBS-3 in Fig. \ref{fig1}),where
$|R\rangle=1/\sqrt{2}(|H\rangle+i|V\rangle)$ and
$|L\rangle=1/\sqrt{2}(|H\rangle-i|V\rangle)$.

%%%%%%%%%%%%%%%%%%%%%%%% Experiment %%%%%%%%%%%%%%%%%%%%%%%%%%%%%%%

\begin{figure}[hbtp]
\includegraphics[width=8.5cm]{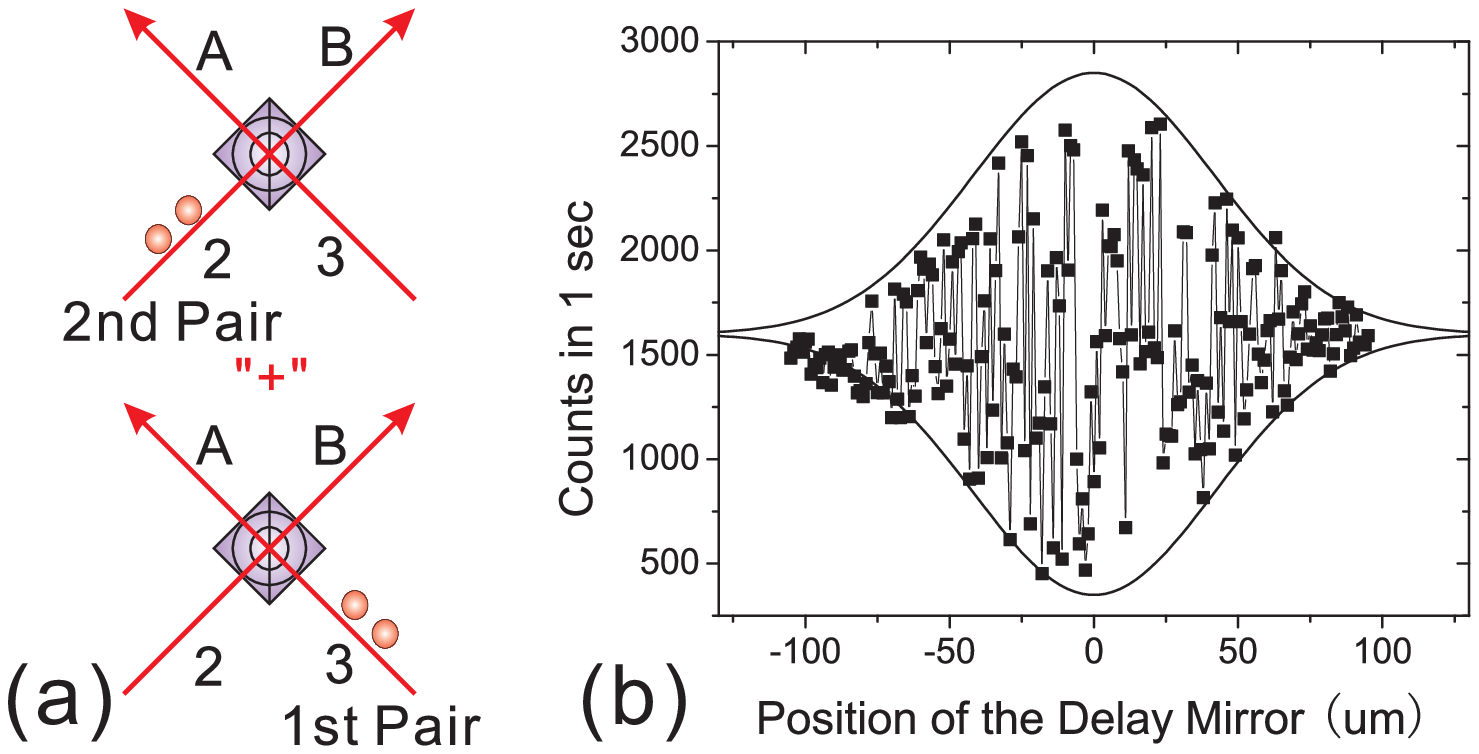}
\caption{(color online). Two-photon Mach-Zehnder interference as a
method to find the overlap on PBS-3. (a) Schematic diagram of the
method. After perfect overlaps on PBS-1 and PBS-2 have been
achieved, by adjusting polarizers and wave plates, two photons
originated from the first pair will be on path 3 with polarization
state of $|1R,1L\rangle_3$ and the other two photons originated from
the second pair will be on path 2 with polarization state of
$|1R,1L\rangle_2$. Consider that the probability of generating two
pairs simultaneously is rarely low, so the coincident click between
detector $D_A$ and detector $D_B$ maybe originate either from the
two photons on path 2 or from the two photons on path 3. So the
two-photon state before PBS-3 can be expressed as
$1/\sqrt{2}(|1R,1L\rangle_3+e^{i\phi}|1R,1L\rangle_2)$ in the case
where these two possibilities interfere. Then passing through PBS-3,
the state will change to
$1/\sqrt{2}(|R\rangle_A|L\rangle_B+e^{i\phi}|L\rangle_A|R\rangle_B)$.
So when we move the delay mirror adjusting the phase $\phi$, the
coincident count between $D_A^H$ and $D_B^H$ will oscillate as a
function of the position. (b) Experimental result of the
oscillation. We can estimate the overlap position from the best fit
of the envelop with Gauss function. \label{overlap}}
\end{figure}

Our scheme works ideally with true single-photon input. But at
present manipulating multi single photons simultaneously is still a
difficult task \cite{Santori2002}. In our proof-of-principle
experiment, we utilized disentangled photons from SPDC
\cite{kwiat1995} sources as the four input photons of the CNOT gate
as shown in Fig. \ref{fig1}(b). Perfect spacial and temporal overlap
on the three PBS are necessary, which is highly related to the
fidelity of the gate. In the experiment narrow-band interference
filters are added in front of each detector to define the exact
spectral emission mode, resulting in a coherence time longer than
the pulse duration. All the photons are collected with single-mode
fibers to define the exact spacial mode. Additional translators are
added in path a1 and a2 to achieve good temporal overlap on PBS-1
and PBS-2. Previously to get the perfect temporal overlap between
photons from different pairs on PBS-3, people have to measure the
four-fold coincident counts as a function of scanning position of
the delay mirror (as shown in Fig. \ref{fig1}(b)). But usually the
four-fold count rate is very low (at the order of 1/sec typically),
which usually makes the scanning process take a long time. In our
experiment, we can overcome this difficulty by utilizing the
two-photon Mach-Zehnder interference effect as shown in Fig.
\ref{overlap}. As a result we can scan two-fold coincident counts
instead, which is much higher than four-fold coincident counts , and
this greatly shortens the process to get photons from different
pairs temporal overlapped, usually we can find the temporal overlap
on PBS-3 in about several tens of seconds only.

%%%%%%%%%%%%%%%%%%%%%%%%%% result %%%%%%%%%%%%%%%%%%%%%%%%%%%%%%%

To evaluate the performance of our gate, first we test the
capability to generate entanglement. We choose the input product
state as $|+\rangle_c|H\rangle_t$. Corresponding to the CNOT
operation, ideally the output state should be $|\Phi^+\rangle_{14}$,
which is a maximal entangled state. To verify this, we measure the
correlation between the polarizations of photon 1 and photon 4, and
the measured visibilities are $(83.8 \pm 5.5)\%$ and $(96.0 \pm
2.8)\%$ for $|H\rangle/|V\rangle$  and $|+\rangle/|-\rangle$ basis,
respectively. As we know for states with a visibility above 71\%,
Bell inequalities \cite{bell1964,chsh1969} can be violated, which is
a important criterion for entanglement.

In order to get the most complete and precise evaluation of a gate,
previously quantum process tomography \cite{poyatos1997} has been
utilized in former experiments
\cite{brien2004,Langford2005,nikolai2005}. However, 256 different
measurement setups are required to evaluate only a CNOT gate. In
contrast, here we utilize a recently proposed method
\cite{hofmann2005} to fully evaluate our gate, in which only 32
measurements are required. From these measurements we can get the
upper and lower bound of the gate fidelity.
\begin{figure}[hbtp]
\includegraphics[width=8.5cm]{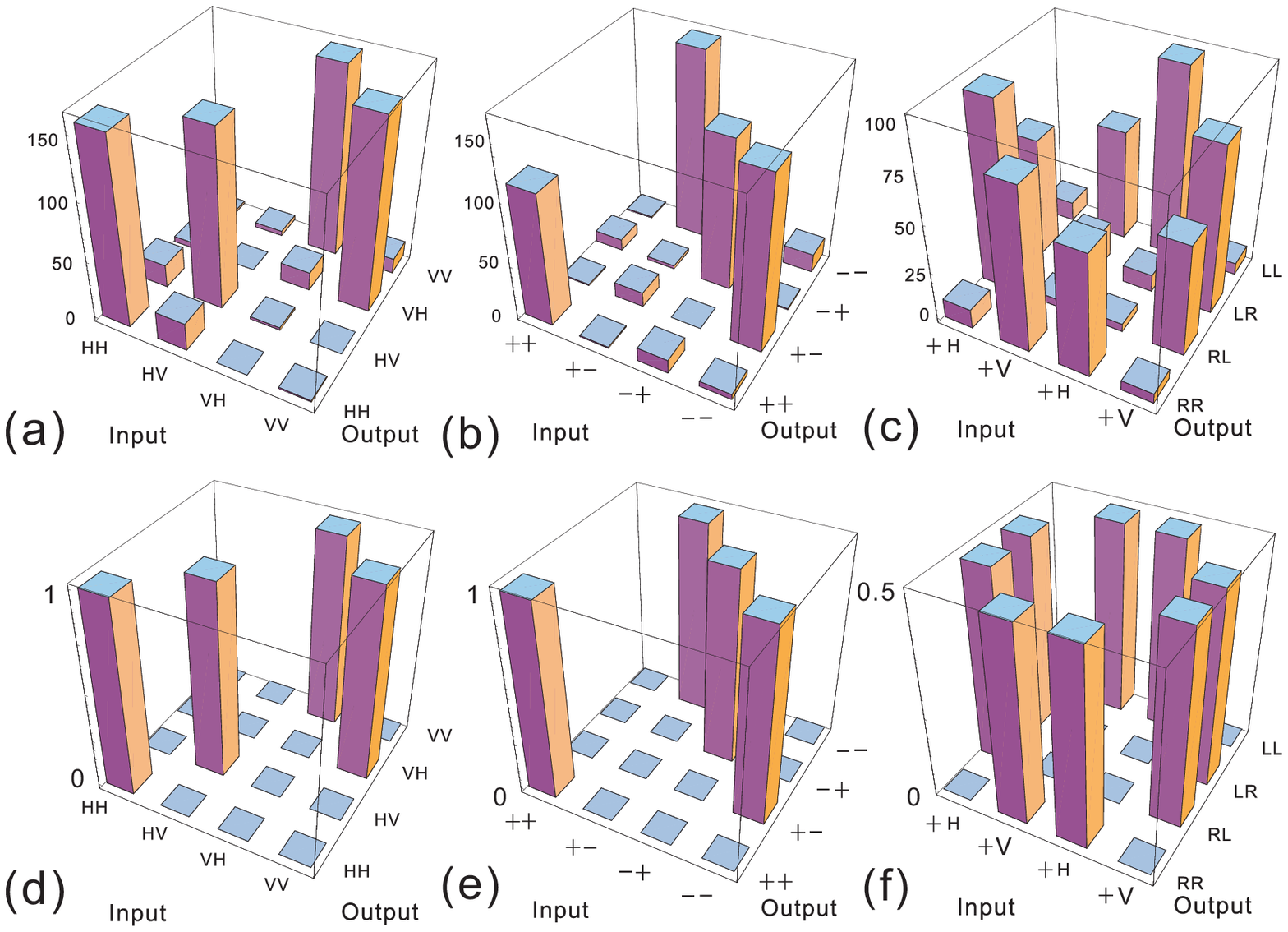}
\caption{(color online). Experimental evaluation of the CNOT gate,
each data point is measured in 320 min for the first three figures.
(a) in the computational basis ($|H\rangle/|V\rangle$). (b) in the
complementary basis ($|+\rangle/|-\rangle$). For (c), the input
control qubit is in the $|+\rangle/|-\rangle$ basis and the input
target qubit is in the $|H\rangle/|V\rangle$ basis, while the output
qubits are measured in the $|R\rangle/|L\rangle$ basis. (d),(e) and
(f) are theoretical values (for the vertical axis, probability is
adopted instead of count rate) for (a), (b) and (c), respectively.
\label{result}}
\end{figure}

As we know, in the computational basis ($|H\rangle/|V\rangle$) under
the CNOT operation, the target qubit flips when the control qubit is
logic 1 (state $|V\rangle$). However, this process gets reversed in
the complementary basis ($|+\rangle/|-\rangle$) that the control
qubit will flip if the target qubit is logic 1 (state $|-\rangle$).
Measurement of the logic functions in these two bases will give a
good estimation of the range of the gate fidelity. The experimental
results are shown in Fig. \ref{result}(a) and Fig. \ref{result}(b).
Let's define the fidelities in these two bases as
\begin{equation}
\begin{array}{rcclcl}
F_{1} & = & 1/4\ [&P(HH|HH)&+&P(HV|HV)
\\
& & \quad +&P(VV|VH)&+&P(VH|VV)\ ]
\\
F_{2} & = & 1/4\ [&P(++|++)&+&P(--|+-)
\\
& & \quad +&P(-+|-+)&+&P(+-|--)\ ] \\
\end{array}
\end{equation}
where each \textit{P} represents the probability to get the
corresponding output state under the specified input state
condition. In order to convert the coincident count rates to
probabilities, we normalize them with the sum of coincidence counts
obtained for the respective input state. In our experiment measured
$F_{1}$ is $(88 \pm 1)\%$ and $F_{2}$ is $(90 \pm 1)\%$. As
discussed in detail in Ref. \cite{hofmann2005}, the upper bound and
low bound of the gate fidelity can be obtained from these two
fidelities as follows:
\begin{equation}
(F_{1} + F_{2} - 1) \leq F \leq \min (F_{1} , F_{2})\,.
\end{equation}
In our experiment the lower and upper bounds of the gate fidelity
are $(78 \pm 2)\%$ and $(88 \pm 1)\%$ respectively. Consider into
the imperfections of the polarizers and waveplates used and the
slightly higher order events (estimated ratio of 6-photon count rate
to 4-photon count rate is only about 0.008), the fidelity of initial
state preparation can be better than 98.9\%. If the initial state
preparation is perfect, the measured gate fidelity will be improved
a little bit. We think that most of the degradation of the fidelity
is due to the imperfection of PBS and the imperfect overlapping on
it.

Recently a new experimental criterion for the evaluation of device
performance has been proposed\cite{hofann2005pra}. It was shown that
a quantum controlled-NOT gate simultaneously performs the logical
functions of three distinct conditional local operations. Each of
these local operations can be verified by measuring a corresponding
truth table of four local inputs and four local outputs.
Specifically, quantum parallelism is achieved if the average
fidelity of the three classical operations exceeds 2/3. As a matter
of fact the fidelity $F_{1}$ and $F_{2}$ are just two of the
required three fidelities. The third fidelity is defined as
\begin{equation}
\begin{array}{ccl}
F_{3}&=&1/4\ [P(RL/+H)+P(LR/+H)
\\
&+&P(RR/+V)+P(LL/+V)+P(RR/-H)
\\
&+&P(LL/-H)+P(RL/-V)+P(LR/-V)]\,.
\end{array}
\end{equation}
The experimental result of $F_{3}$ is shown in Fig. \ref{result}(c)
with the measured value $(90 \pm 1)\%$. The average fidelity of
$F_{1}$, $F_{2}$ and $F_{3}$ is $(89 \pm 1)\%$, exceeding the
boundary 2/3, which shows that quantum parallelism of our CNOT gate
has been achieved and the performance of the gate can not be
reproduced by local operations and classical communications.

%%%%%%%%%%%%%%%%%%%%%%%%%%% Conclusion %%%%%%%%%%%%%%%%%%%%%%%%%%%%%

In summary, we have presented and experimentally demonstrated a
novel scheme to realize the optical nondestructive CNOT gate without
using entangled photons but only single photons instead. With much
fewer measurements compared with quantum process tomography
\cite{poyatos1997}, we got a good estimation of the gate fidelity
(between $(78 \pm 2)\%$ and $(88 \pm 1)\%$), showing a great
improvement compared with previous experiments (In \cite{sara2004}
severe noise from unwanted two-pair events has been subtracted from
the experiment result; in \cite{zhao2005} five photons were involved
to avoid two-pair events, resulting in rather low visibility).
Moreover, we have also shown that quantum parallelism was achieved
in our CNOT gate. We believe that our experiment and the methods
developed in this experiment would have various novel applications
in the fields of both linear optical quantum information processing
and quantum communication with photons.

%%%%%%%%%%%%%%%%%%%%%%%% Acknowledgements %%%%%%%%%%%%%%%%%%%%%%%%%

\begin{acknowledgments}
This work is supported by the NNSF of China, the CAS, and the
National Fundamental Research Program (under Grant No.
2006CB921900).
\end{acknowledgments}

\end{document}